\documentclass[10pt,aps,prd,onecolumn,showpacs,amsmath,amssymb,nofootinbib,eqsecnum,preprintnumbers,superscriptaddress]{revtex4-2}

\usepackage{mathtools}
\usepackage{amsmath}
\usepackage{subcaption}
\usepackage{caption}
\usepackage{xcolor}
\usepackage{graphicx}
\usepackage{mathrsfs}
\usepackage{tikz}
\usepackage{comment}
\usepackage[colorlinks=true,
            citecolor=red,
            linkcolor=blue,
            urlcolor=violet,
            filecolor=cyan,
            backref=false]{hyperref}
\usepackage{multirow}

\newcommand\feq{\mathrel{\phantom{=}}}

\begin{document}

\title{Carroll black holes in (A)dS and their higher-derivative modifications}

\author{Poula Tadros}
\email{poula.tadros@matfyz.cuni.cz}
\affiliation{Institute of Theoretical Physics, Faculty of Mathematics and Physics, Charles University,
V Hole\v{s}ovi\v{c}k\'ach 2, Prague 180 00, Czech Republic}

\author{Ivan Kol\'a\v{r}}
\email{ivan.kolar@matfyz.cuni.cz}
\affiliation{Institute of Theoretical Physics, Faculty of Mathematics and Physics, Charles University,
V Hole\v{s}ovi\v{c}k\'ach 2, Prague 180 00, Czech Republic}

\date{\today}

\begin{abstract}
We define the Carrollian black holes corresponding to the limit of Schwarzschild-(A)dS spacetime and its higher-derivative counterpart known as Schwarzschild-Bach-(A)dS spacetime, which is also a static spherically symmetric vacuum solution of quadratic gravity. By analyzing motion of massive particles in these geometries, we found that: In the case of Schwarzschild-(A)dS, a (nearly) tangential particle from infinity will wind around the extremal surface with a finite number of windings depending on the impact parameter and the cosmological constant. In Schwarzschild-Bach-(A)dS, a particle passing close enough to the extremal surface will have an infinite number of windings; hence, it will not escape to asymptotic infinity as in Schwarzschild-(A)dS. We also calculate the thermodynamical quantities for such black holes and argue that it is analogous to an incompressible thermodynamical system with divergent entropy when the temperature goes to zero (in the strict Carroll limit). We then define a divergent specific heat that can be positive, negative, or zero.
\end{abstract}

\maketitle

\section{Introduction}
\textit{Quadratic gravity} is a modification of \textit{general relativity (GR)} containing quadratic terms in curvature in the Lagrangian. It can be derived as an effective field theory by truncating the bosonic sector of string theory to the next-to-leading order (with the leading order being GR) \cite{ZUMINO1986109,ZWIEBACH1985315,Forger:1996vj,Myers:1987yn,Alvarez-Gaume:2015rwa}. Alternatively, it can also be achieved by imposing a maximal momentum in string theory \cite{Nenmeli:2021orl}. By adding quadratic curvature terms in the Lagrangian, the theory becomes renormalizable, but possesses extra ghost degrees of freedom \cite{Stelle:1976gc}. The Lagrangian for quadratic gravity is given by
\begin{equation}\label{action}
  \mathcal{L}= \gamma (R-2\Lambda) + \alpha R_{\mu\nu}R^{\mu\nu} + \beta R^2,  
\end{equation}
where $R_{\mu\nu}$ is the Ricci tensor, $R$ is the Ricci scalar, $\Lambda$ is the cosmological constant and $\alpha,\beta,\gamma$ are parameters of the theory. The equations of motion are given by  
\begin{equation}
    \gamma (R_{\mu\nu}-\tfrac{1}{2}Rg_{\mu\nu}+ \Lambda g_{\mu\nu}) - 4\alpha B_{\mu\nu} + 2\beta (R_{\mu\nu} - \tfrac{1}{4}Rg_{\mu\nu} + g_{\mu\nu}\Box - \nabla_{\mu}\nabla_{\nu})R=0,
\end{equation}
where $\Box = g^{\mu\nu}\nabla_{\mu}\nabla_{\nu}$ and $B_{\mu\nu} = (\nabla^{\sigma}\nabla^{\rho} + \tfrac{1}{2}R^{\sigma \rho})C_{\mu\sigma\nu\rho}$ is the Bach tensor with $C_{\mu\sigma\nu\rho}$ being the Weyl tensor.

\noindent This theory has the Schwarzschild metric as a solution; however, it also contains a one-parameter family of static spherically symmetric solutions called the \textit{Schwarzschild-Bach} black holes whose deviation from Schwarzschild is measured by a quantity called the \textit{Bach parameter} denoted by $\delta$  \cite{Lu:2015cqa,Lu:2015psa,Podolsky:2019gro,Pravdova:2023nbo}. Such black holes are static spherically symmetric solutions of the field equations. Thus, their metric is given by the form 
\begin{equation}\label{1}
    ds^2 = -h(r)dt^2+ \tfrac{dr^2}{f(r)} + r^2 d\Omega^2,
\end{equation}
where $d\Omega^2 = d\theta^2 + \textrm{sin}(\theta)d\phi^2$, $h(r)$ and $f(r)$ are functions to be deduced from the field equations.
The solutions to the field equations by substituting with this metric can not be given analytically due to the complexity and non-linearity of the field equations, however, using Frobenius analysis, solutions are given in the form of infinite series for the functions for $h$ and $f$ near the horizon depending on the specific solution to be considered i.e.
\begin{equation}
    \begin{aligned}
       f(r) = \sum_{n=0}^{\infty}f_{\omega+n}(r-r_0)^{\omega + n}, \ \ \  h(r) = h_t\bigg((r-r_0)^t +\sum_{m=1}^{\infty}h_{t+n}(r-r_0)^{t+n}\bigg),
    \end{aligned}
\end{equation}
where $\omega$ and $t$ are parameters in the Frobenius series that will be used to differentiate different classes of solutions. As shown in \cite{Lu:2015psa}, there are three classes of Frobenius solutions:
\begin{enumerate} \itemsep0em
    \item The class defined by setting $\omega = t = 0$, denoted by $(0,0)$, which has no horizon.
    \item The class defined by setting $\omega = 1$ and $t=0$, denoted by $(1,0)$, which defines a wormhole.
    \item the The class defined by setting $\omega = t = 1$, denoted by $(1,1)$, which contains the Schwarzschild solution as well as the Schwarzschild-Bach solutions.
\end{enumerate}
 For more details on the aforementioned classes and solutions, we refer the reader to \cite{Svarc:2018coe}. There are also other classes derived from non-Frobenius series, but we will not discuss those. In this paper, we are only interested in the class $(1,1)$ since it contains the class of Schwarzschild-Bach black holes. In this work, we will construct the Carrollian versions of these black holes. 

The \textit{Carroll group} is defined as the ultra-local limit of the Poincar\'e group where the speed of light $c$ goes to zero, $c\to 0$. It was defined independently by Levy-Leblond \cite{Levy-Leblond} and Sen Gupta \cite{Gupta} in the 1960s as the In\"on\"u-Wigner contraction of the Poincar\'e group. Afterwards, it was recognised that this group appears in near black hole horizon physics \cite{Donnay:2019jiz,Penna:2018gfx,Grumiller:2018scv} and at null infinity \cite{Penna:2015gza,Duval:2014uva,Barnich:2009se,Barnich:2011ty,Barnich:2011mi}. Since then, it has been studied from various perspectives and applied in various areas of physics \cite{Zhang:2023jbi,Bergshoeff_2014,Marsot:2021tvq,Marsot:2022imf,Bagchi:2022eui,Kubakaddi_2021,Kononov_2021,PhysRevD.106.085004,Chen:2023pqf,Bergshoeff:2022eog,Henneaux:2021yzg,Bagchi:2019xfx,Bagchi:2019clu,Bagchi:2021gai,PhysRevD.103.105001,Bagchi:2023ysc,Ciambelli:2018wre,Ciambelli:2018xat,campoleoni2019two,Ciambelli:2020eba,10.21468/SciPostPhys.9.2.018,PhysRevLett.123.111601,Bagchi:2021ban,Cardona:2016ytk,Perez:2021abf,Perez:2022jpr,Hartong:2015xda,Figueroa-OFarrill:2021sxz,Hansen:2021fxi,Gomis:2020wxp,Bergshoeff:2022qkx,Guerrieri:2021cdz,Hansen:2020wqw,Herfray:2021qmp,Chandrasekaran:2021hxc,Bagchi:2019clu,Ciambelli:2019lap,Bicak:2023vxs,Marsot:2022qkx, Kasikci:2023tvs,deBoer:2021jej}. 

Despite all the work in this direction \textit{Carroll black holes}, which were first properly introduced in \cite{Ecker:2023uwm} were rarely studied (for example in \cite{Ciambelli:2023tzb}). They are defined on a Carrollian manifold, which is a manifold endowed with a nowhere vanishing vector field $v^{\mu}$ and a degenerate metric $h_{\mu\nu}$ such that $v^{\mu}h_{\mu\nu}=0$. Furthermore, we define an associated form $E_{\mu}$ to $v^{\mu}$ such that $E_{\mu}v^{\mu}=1$. Notice that $E_{\mu}$ plays the role of an Ehresmann connection and need not be covariantly constant. A Carrollian manifold can alternatively be defined by contracting the light cone at each point of a Lorentzian manifold $\mathcal{M}$. This splits its tangent bundle $T\mathcal{M}$ into a vertical bundle $\mathrm{Ver} \mathcal{M}$ (a vector bundle whose fibers are isomorphic to the timelike vector flow) and horizontal bundles $\mathrm{Hor} \mathcal{M}$ (whose fibers are isomorphic to spacelike slices of $\mathcal{M}$), i.e, $T\mathcal{M} = \mathrm{Ver} \mathcal{M} \oplus \mathrm{Hor} \mathcal{M}$. This results in a foliation of the manifold, by lines corresponding to the flow of the kernel of the induced metric i.e. $v^{\mu}$, into codimension 1 submanifolds with metric $h_{\mu\nu}$ as leaves (as reminiscents of the spacelike slices in the original Lorentzian manifold). We note that is different from the foliation of Lorentzian manifolds into codimension 1 submanifolds without the light cone contraction since the timelike vector flow is not restricted from one leave to the other. We denote the adapted coordinates in the leaves by $x_{\perp}^{\mu}$, and the coordinate in direction of $v^{\mu}$ by $x^{\mu}_{\parallel}$. A detailed discussion of Carrollian geometry and Carrollian symmetries can be found in \cite{deBoer:2023fnj,Borthwick:2024wfn,Ciambelli:2023xqk}. Carrollian manifolds can be endowed with differential structures that require both $v^{\mu}$ and $h_{\mu\nu}$ to be covariantly constant. This defines special torsional connections; see \cite{Hansen:2021fxi} for more details. A Carroll black hole defined as a solution of a Carroll gravity theory (a gravity theory defined on a Carroll manifold) that has a Carroll extremal surface (a surface for which the norm of $v^{\mu}$ is zero) and has certain thermodynamical properties stated in \cite{Ecker:2023uwm}.

In this paper, we study Carroll black holes further as geometries and as thermodynamical systems. We introduce the Carrollian version of  Schwarzschild-(A)dS and Schwarzschild-Bach-(A)dS black holes as solutions to the Carroll limit of GR and quadratic gravity and analyze the geodesic motion on their geometries. We show that, in the case of Carroll Schwarzschild-(A)dS, a near-tangential particle winds around the extremal surface with a \textit{finite} number of windings, similar to the Carroll Schwarzschild black hole, but this number receives \textit{corrections} due to the cosmological constant. On the other hand, in the case of Schwarzschild-Bach-(A)dS black hole, the number of windings is \textit{infinite} regardless of the cosmological constant, i.e. this feature is only due to the higher-curvature terms added to the Lagrangian. We also study the thermodynamical properties of such black holes and compare them to the incompressible thermodynamical system with a divergent entropy and vanishing temperature. In what follows, we introduce a scale for each dimensionful quantity and work with the dimensionless quantities obtained by dividing by these scales. The paper is organized as follows:
\begin{itemize} \itemsep0em 
    \item In Sec. \ref{section2}, we introduce Carroll Schwarzschild-Bach black holes and deduce its geodesics using the method in \cite{Ciambelli:2023tzb}. We then derive the circularity condition and show that nearly tangential particles wind an infinite number of times around the extremal surface. Then we compute its thermodynamical quantities (entropy, temperature and energy) and show that its specific heat is always negative.
    \item In Sec. \ref{section 3}, we perform a similar analysis for Carroll Schwarzschild-(A)dS black holes. Specifically, we calculate the geodesics, circularity conditions, the number of windings, which is finite, for nearly tangential particles as a function of the impact parameter, and also study its thermodynamics. In this case we show that the specific heat can be positive or negative depending on the radius of the extremal surface and the cosmological constant.
    \item In Sec. \ref{section 4}, we move on to Carroll Schwarzschild-Bach-(A)dS black holes. We closely follow the previous derivations from Sec. \ref{section2} but with non zero cosmological constant, and show that the heat capacity can be positive or negative. 
    \item In Sec. \ref{sec 8}, we conclude the paper with a brief summary and possible future directions.
\end{itemize}

\section{Carroll Schwarzschild-Bach black hole}\label{section2}

\subsection{Geodesics}\label{sec 2}
We start by computing the Carroll limit of Schwarzschild-Bach black hole and then derive the geodesic equations for a test particle in that geometry following the procedure in \cite{Ciambelli:2023tzb}. The metric for the Schwarzschild-Bach black hole is \eqref{1}, its Carroll limit is derived by restoring the factors of $c$ so it can be written as
\begin{equation}
ds^2 = -h(r)c^2 dt^2+ \tfrac{dr^2}{f(r)} + r^2 d\Omega^2,    
\end{equation}
then taking the limit $c \to 0$. The result is a degenerate metric, and a vector in its kernel (along with its associated one form) given by 
\begin{equation}\label{Carroll Schwarzschild Back black hole with zero cosmological constant}
    \begin{aligned}
 h_{\mu\nu}dx^{\mu}dx^{\nu} =  \tfrac{dr^2}{f(r)} +r^2d\Omega^2, & \ \ \  
       v^{\mu}\partial_{\mu} = \tfrac{-1}{h(r)}\partial_t,& E_{\mu}dx^{\mu} = -h(r)dt
    \end{aligned}
\end{equation}
Notice that due to the fact that the Lorentzian metric is static, the Carroll limit looks like a partition of the Lorentzian data. However, one can easily check that $h_{\mu\nu}v^{\mu}=0$ and $E_{\mu}v^{\mu}=1$ which demonstrate the fact that it is indeed a Carrollian construction rather than just a Lorentzian data partition. By straightforward calculations, we can also verify that \eqref{Carroll Schwarzschild Back black hole with zero cosmological constant} satisfies the Hamiltonian constraints and the magnetic equations of motion of the theory labeled (2,4) in \cite{Tadros:2023teq}. That is a Carrollian theory derived from the limit of quadratic gravity where the parameters of the action \eqref{action} are assumed to be of the form $\alpha = c^2 \alpha'$ and $\beta = c^4 \beta'$, where $\alpha'$ and $\beta'$ have no $c$ dependencies. This means that the chosen foliation gives a solution for the Carrollian theory, and since it has a Carroll extremal surface corresponding to $f(r)=0$ (where the degenerate metric diverges, giving rise to a structural singularity \cite{Ecker:2023uwm}), this solution can be called a Carroll black hole whose metric is defined near the extremal surface.  Thus, we
expand $f(r)$ into powers of $r-r_0$ as follows:
\begin{equation}
    f(r) = f_1 (r-r_0) + f_2 (r-r_0)^2 + \cdots,
\end{equation}
where $f_1 = \tfrac{1+\delta}{r_0}$, $f_2 = \tfrac{1}{r_0(1+\delta)}1+3\delta (1-\tfrac{r_0^2}{8k}) + 2\delta^2$ and $\delta$ is the Bach parameter.

 To study the dynamics of particles, an action needs to be constructed for geodesic particles in Carrollian geometry. This was done in \cite{Ciambelli:2023tzb}, first by proving that the only Carroll invariant quantities with two derivatives or less are : $h_{\mu\nu}\dot{x}_{\perp}^{\mu}\dot{x}_{\perp}^{\nu}$ 
 , $\dot{x}_{\parallel}$ 
 as well as a constant. Thus, the most general action consisting of these quantities, and the einbein of the worldline $e$, is given by a linear combination of the three mentioned quantities, that is,
\begin{equation}\label{geodesic action}
    I = \int d\tau e (g_0 + g_1e^{-2}h_{\mu\nu}\dot{x}^{\mu}_{\perp}\dot{x}_{\perp}^{\nu} + g_2e^{-2}\dot{x}^2_{\parallel}),
\end{equation}
where $\dot{x}=\tfrac{dx}{d\tau}$. This action is Carroll diffeomorphism invariant since only Carroll diffeomorphism invariant quantities couple to $\dot{x}^{\mu}$. Notice that we do not require local Carroll boost symmetry which forbids local energy flux and in turn forbids the motion of massive particles. In our case such symmetry is broken so we expect non trivial dynamics. In this paper, we choose $x^{\mu}_{\perp}=r,\theta,\phi$ to be the spherical coordinates on the spatial slice, $x^{\mu}_{\parallel}=t$ is the time coordinate (normal to the spatial slice), and $g_0,g_1,g_2$ are arbitrary constants. Without loss of generality, we assume that the motion is in the equatorial plane $\theta = \tfrac{\pi}{2}$. Substituting \eqref{Carroll Schwarzschild Back black hole with zero cosmological constant} into \eqref{geodesic action}, we get
\begin{equation}\label{geodesic action in coord}
    I = \int d\tau e (g_0 + g_1 e^{-2}(\tfrac{\dot{r}^2}{f(r)}+r^2\dot{\phi}^2) + g_2e^{-2}\dot{t}^2).
\end{equation}

\noindent The equation for $t(\tau)$ is given by
\begin{equation}\label{teqn}
    \ddot{t}=0 \implies t=F\tau + t_0,
\end{equation}
where $F$ and $t_0$ are arbitrary constants.
The equation for $\phi(\tau)$ is
\begin{equation}\label{phieqn}
    \partial_{\tau}(r^2\dot{\phi})=0 \implies \dot{\phi}=\tfrac{l}{r^2},
\end{equation}
where $l$ is an arbitrary constant.

 Using the above equations, we can now derive the constraint for this action due to reparametrization invariance by varying the action with respect to $e$, the result is
\begin{equation}
    e^2 = \tfrac{g_1}{g_0}(\tfrac{\dot{r}^2}{f(r)}+r^2\dot{\phi}^2)+\tfrac{g_2}{g_0}\dot{t}^2,
\end{equation}
substituting from \eqref{teqn} and \eqref{phieqn}, we end up with
\begin{equation}
    e^2 = \tfrac{g_1}{g_0}(\tfrac{\dot{r}^2}{f(r)}+l^2)+\tfrac{g_2}{g_0}F^2.
\end{equation}
Upon inserting in \eqref{geodesic action in coord}, we get
\begin{equation}
    I=  \int 2 \sqrt{g_1(\tfrac{\dot{r}^2}{f(r)}+l^2)+g_2 F^2}d\tau.
\end{equation}
 
Following \cite{Ciambelli:2023tzb}, the definition of the particle's energy is chosen to be
\begin{equation}
    E= \tfrac{g_0}{2g_1}e^2 - \tfrac{g_2}{2g_1}F^2.
\end{equation}
 The presence of the coefficients in the previous formulae indicates that not all Carrollian formulae can be derived as a limit of a Lorentzian one. Indeed, the formulae derived from the Carroll limit of the corresponding Lorentzian formulae must have fixed ratios of coefficients rather than arbitrary, namely, $\tfrac{g_1}{g_2} = \mathcal{O}(c^{-2})$ which implies that one can only have only the electric or the magnetic terms in the action, and the following formulae as proven in \cite{deBoer:2023fnj}. Here, we consider Carroll diffeomorphism actions without referring to a parent Lorentzian manifold. This allows us to add Carroll diffeomorphism invariant terms with arbitrary coupling constants without worrying about the Lorentz invariance of any Lorentzian manifold. Hence having interactions between the electric and the magnetic terms in a consistent way giving rise to more interesting dynamics as we will see below.

 Using this definition, the geodesic equation is written as
\begin{equation}\label{geodesic equation}
    \tfrac{\dot{r}^2}{2}+V^{\textrm{eff}}(r)=E,
\end{equation}
where $V^{\textrm{eff}}(r)=\tfrac{l^2f(r)}{2r^2}-(f(r)-1)E$ is the effective potential. Notice that when $f(r)=1-\tfrac{2M}{r}$, we recover the results for Schwarzschild derived in \cite{Ciambelli:2023tzb}.
 Now, we write the circularity conditions as
\begin{equation}
    \begin{aligned}
        V^{\textrm{eff}}=E, & \hspace{10pt} \tfrac{dV^{\textrm{eff}}}{dr}=0,
    \end{aligned}
\end{equation}
the first condition ensures that the geodesics are circles i.e. $\dot{r}=0$, and the second condition is to guarantee that the geodesic is equipotential i.e. there is no force generated by the effective potential at any point of the circular trajectory.
The conditions, in our case, translate to
\begin{equation}\label{circularity conditions}
    \begin{aligned}
       f(r)(\tfrac{1}{b^2}-\tfrac{1}{r^2})l^2&=0,\\
    (\tfrac{1}{b^2}-\tfrac{1}{r^2})l^2f'(r)+\tfrac{l^2}{r^2}f(r)&=0,  
    \end{aligned}
\end{equation}
where we introduced the impact parameter $b=\tfrac{l}{\sqrt{2E}}$. \footnote{We call it impact parameter inspired by the Lorentzian Schwarzschild case where it is the solution for $\lim_{r\to\infty}|\dot{r}|^2=\tfrac{l}{b}$. It will become evident later that this naming is consistent.} The first condition is satisfied at $r=b$ or $f(r)=0$. If $r=b$, then the second condition yields $f(r)=0$, i.e. at $r=r_0$, that is, circularity occurs only on the extremal surface. Similarly if $f(r)=0$, i.e. $r=r_0$, then we have either $r=b$ or $f'(r)=0$. The latter results in a contradiction since when expanding 
\begin{equation}
    f(r) = f_1(r-r_0) + f_2 (r-r_0)^2 + \dots,
\end{equation}
the condition $f'(r_0)=0$ implies $f_1 = 0$ which is not consistent with the field equations for quadratic gravity \cite{Lu:2015cqa}. Thus, the circularity condition is
\begin{equation}\label{circularity conditions 1}
\begin{aligned}
    b=r_0.
    \end{aligned}
\end{equation}

 To study the stability of the orbits, we compute the second derivative $\tfrac{d^2V^{\textrm{eff}}}{dr^2}$, we get
\begin{equation}
 \tfrac{d^2V^{\textrm{eff}}}{dr^2} = (\tfrac{l^2}{2r^2}-E)f''(r)- \tfrac{2l^2}{r^3}f'(r) + \tfrac{3l^2}{r^4}f(r).   
\end{equation}
Employing the circularity conditions \eqref{circularity conditions 1}, we obtain
\begin{equation}
    \tfrac{d^2V^{\textrm{eff}}}{dr^2}(r_0) = - \tfrac{2l^2}{r_0^3}f'(r_0).
\end{equation}
In the case $f(r)=1-\tfrac{2M}{r}$, this quantity is negative at the horizon, and we do not have stable orbits which agrees with the conclusion in \cite{Ciambelli:2023tzb}. Expanding $f(r)$ near the horizon and setting $r=r_0$, we get
\begin{equation}\label{second derivative}
\tfrac{d^2V^{\textrm{eff}}}{dr^2}(r_0) = - \tfrac{2l^2}{r_0^3}f_1.
\end{equation}
Following \cite{Lu:2015cqa}, we can write $f_1 
= \tfrac{1+\delta}{r_0}$. Substituting into \eqref{second derivative}, we have
\begin{equation}
    \tfrac{d^2V^{\textrm{eff}}}{dr^2}(r_0) = - \tfrac{2l^2}{r_0^4}(1+\delta).
\end{equation}
 There are three cases depending on the value of the Bach parameter:
\begin{enumerate} \itemsep0em 
 \item $\delta > -1$:  $ \tfrac{d^2V^{\textrm{eff}}}{dr^2}(r_0)$ is negative and there are no stable orbits. This class includes Schwarzschild ($\delta$ = 0).
 \item $\delta < -1$: $ \tfrac{d^2V^{\textrm{eff}}}{dr^2}(r_0)$ is positive and there are stable orbits.
 \item $\delta = -1$: $f_1 = 0$, which is incompatible with the field equations as mentioned above.
\end{enumerate}
\noindent Thus, Carroll Schwarzschild-Bach black holes can have stable circular orbits on the extremal surface, unlike the Carroll Schwarzschild case. 

\begin{figure}[htb]
    \centering
    \begin{subfigure}[t]{\textwidth}
        \centering
        \includegraphics[width=0.5\linewidth]{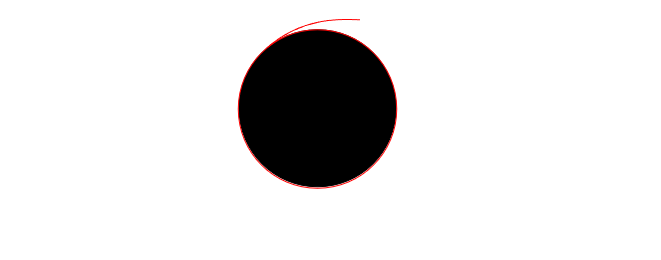}%
        \hfill
        \includegraphics[width=0.5\linewidth]{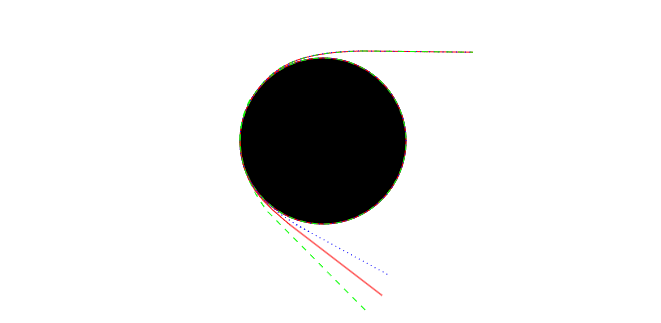}
    \end{subfigure}
    \caption{In the case of Carroll Schwarzschild-Bach black hole (the left picture), in the region near the extremal surface i.e. where the black hole's metric is defined, a particle will undergo an infinite number of windings around the extremal surface for any impact parameter $b \approx r_0$. The trajectory takes the shape of an infinitely winding spiral which infinitesimally approaches the extremal surface but never touches it. That is in contrast with Carroll Schwarzschild-(A)dS (the right picture where the green lines represent the trajectory of a particle near Carroll Schwarzschild-dS black hole, the red lines represent the trajectory near Carroll Schwarzschild, and the blue lines represent the trajectory near Carroll-Schwarzschild-AdS black hole) where the winding number is finite but differ depending on the value of the cosmological constant if we fix all the other variables. In this case, a positive cosmological constant, the number of windings decrease in comparison to Carroll Schwarzschild (the particle gets ejected earlier), while in the case of a negative cosmological constant the number of windings increase (the particle gets ejected later).}
    \label{fig:enter-label1}
\end{figure}

Now we calculate the deflection angle for a geodesic particle. From the previous equations, we can deduce that
\begin{equation}
    \frac{d\phi}{dr}=\frac{b}{r\sqrt{r^2-b^2}}\frac{1}{\sqrt{f(r)}}.
\end{equation}
Expanding $f(r)$ near the extremal surface, we obtain
\begin{equation}
   \frac{d\phi}{dr}= \frac{b}{r\sqrt{r^2-b^2}}\bigg( \frac{1}{\sqrt{f_1 (r-r_0)}} - \frac{f_2}{2f_1^{3/2}}\sqrt{r-r_0} \bigg).
\end{equation}
Defining $u=\frac{1}{r}$ and $u_0 = \frac{1}{r_0}$, we can write
\begin{equation}
    \frac{d\phi}{du} = -\frac{b}{\sqrt{1-u^2b^2}}\bigg( \frac{\sqrt{u}}{\sqrt{f_1}} \frac{1}{\sqrt{1-\frac{u}{u_0}}} - \frac{f_2}{2\sqrt{u}f_1^{3/2}}\sqrt{1-\frac{u}{u_0}}  \bigg).
\end{equation}
The deflection angle of a geodesic particle with arbitrary starting and ending points (both points need to be near the horizon since this metric is only valid in this region) is calculated by integrating with respect to $u$. The result is
\begin{equation} \label{deflection angle} 
    \begin{aligned}
 \phi &= \frac{1}{(1 +\delta)^{3/2} \sqrt{\frac{b+\frac{1}{u}}{b}} \sqrt{1-b^2 u^2} \sqrt{u_0-u}} \bigg(b u^{3/2} \sqrt{\frac{1-\frac{u_0}{u}}{b u_0+1}} \bigg(\sqrt{\frac{b+\frac{1}{u}}{b}} \sqrt{1-\frac{1}{b^2 u^2}} (2 (1 +\delta)   + \frac{1}{(1 +\delta)}[1 + \\& \delta (3 - \tfrac{3}{8u_0^2k})+  2\delta^2]) \Pi \bigg(2;\sin ^{-1}\bigg(\frac{\sqrt{\frac{b+\frac{1}{u}}{b}}}{\sqrt{2}}\bigg)|\frac{2 b u_0}{b u_0+1}\bigg)
        +\frac{u_0}{(1 +\delta)}  [1 + \delta (3 - \tfrac{3}{8u_0^2k})+2\delta^2] \bigg(b+\frac{1}{u}\bigg) \sqrt{1-\frac{1}{b u}} \\& \times F\bigg(\sin ^{-1}\bigg(\frac{\sqrt{\frac{b+\frac{1}{u}}{b}}}{\sqrt{2}}\bigg)|\frac{2 b u_0}{b u+1}\bigg)\bigg)\bigg). 
    \end{aligned}
\end{equation}
where $\Pi(n;x|m)$ is the incomplete elliptic integral of the third kind and $F(x|n)$ is the elliptic integral of the first kind. We should also take the limit $b \to 1/u_0$ to ensure that the particle enters the area where the solution is valid. To obtain the deflection angle of the particle, we evaluate this function at a point and take the difference to the value of the function at the turning point $u= 1/b$. The problem with this procedure is that \eqref{deflection angle} diverges at $u = 1/b$. In an attempt to get a finite value from the above formula, we introduce a regularization parameter $a$ so that the function is evaluated at $1/b - a$. 
 The result diverges in the limit $a \to 0$ for any choice of the parameters in the theory, namely $k$ and $\delta$. This is mainly due to the divergence of $F(x|n)$ and $\Pi(n;x|n)$ in the limits $x \to 1$ and $n \to 1$, which are precisely the limits $a \to 0$ and $b \to u_0$, necessary for the consistency of the calculations. Thus, we can conclude that the particle undergoes an infinite number of windings (since the deflection angle is infinite) as shown in Fig. \ref{fig:enter-label1}, i.e. the addition of higher derivative terms to the action results in the situation where particles are trapped near the extremal surface if they come close enough. This can be compared to Carroll Schwarzschild's case where a particle winds a finite number of times and then gets ejected to asymptotic infinity as shown in Fig. \ref{fig:enter-label1}. 


\subsection{Thermodynamics}\label{sec.3}
Let us now analyze the thermodynamics of the Carroll Schwarzschild-Bach black hole. The Carroll black hole \eqref{Carroll Schwarzschild Back black hole with zero cosmological constant} can be thought of as a solution of the magnetic limit of theory (2,4) in \cite{Tadros:2023teq}. This is a Carrollian theory obtained from the Carroll limit of quadratic gravity where the parameters of the action \eqref{action} are considered of the form $\alpha = c^4\alpha'$ and $\beta = c^2 \beta'$ such that $\alpha'$ and $\beta'$ are free from $c$ dependencies. Using Wald's formalism \cite{Wald:1993nt,Iyer:1994ys}, we compute the entropy in two different ways:

\begin{enumerate}
    \item Calculating the entropy of the relativistic black hole, then taking the Carroll limit.
    \item Calculating the entropy from the Carrollian Lagrangian for the Carroll black hole.
\end{enumerate}
We will show that they give the same entropy. The relativistic action that contributes to the magnetic limit of (2,4) (see \cite{Tadros:2023teq} for more details) is given by 
\begin{equation}\label{3}
    S=\int_M d^4x \sqrt{-g}(\tfrac{c^3}{16\pi G}R+ \beta R^2 - \alpha R^{\mu\nu}R_{\mu\nu}).
\end{equation}
Using Wald's formula i.e.
\begin{equation}
    \mathscr{S} = -2\pi \int_H \epsilon_{\mu\nu}\epsilon_{\sigma \rho}\bigg[\tfrac{\partial \mathcal{L}}{\partial R_{\mu\nu\sigma\rho}} + \nabla_{a_1}\tfrac{\partial \mathcal{L}}{\partial \nabla_{a_1} R_{\mu\nu\sigma\rho}} + \cdots + \nabla_{a_1}\nabla_{a_2}\cdots\nabla_{a_n}\tfrac{\partial \mathcal{L}}{\partial \nabla_{a_1}\nabla_{a_2}\cdots\nabla_{a_n} R_{\mu\nu\sigma\rho}} \bigg]\sqrt{g}d^nx,
\end{equation}
where the scripted $\mathscr{S}$ denotes the entropy, the integral is over the horizon, $h$ is the determinant of the induced metric on the horizon, $\mathcal{L}$ is the Lagrangian of the gravitational theory and $\epsilon_{\mu\nu}$ is the binormal to the horizon satisfying $\epsilon_{\mu\nu}\epsilon^{\mu\nu}=-2$. Using this formula together with the theory \eqref{3}, the entropy is 
\begin{equation}
    \mathscr{S}= \tfrac{A_H c^3 k}{4 \hbar G} - \tfrac{2 \pi \alpha k}{\hbar}\int_H \epsilon_{\mu\nu}R^{\mu\nu} \sqrt{h}d^2x,
\end{equation}
where $A_H$ is the area of the horizon, $k$ is the Boltzmann constant, $h$ is the determinant of the metric on the horizon. Using the equations of motion for quadratic gravity, and the fact that Schwarzschild-Bach black hole is spherically symmetric, we can perform the integral to get the result
\begin{equation} 
    \mathscr{S} = \tfrac{A_H c^3 k}{4 \hbar G} - \tfrac{A_H \pi c^2 \delta \alpha'}{r_0^2}.
\end{equation}
So far this is the result for the Lorentzian black hole which matches with the result in \cite{Svarc:2018coe} up to rescalings of the parameters. In the Carroll limit, the Newton's constant is rescaled to be $G_M = c^{-4}G$ since the Carroll limit is considered the strong gravity limit \cite{Anderson:2002zn}. The result is then
\begin{equation}\label{entropy}
    \begin{aligned}
        \mathscr{S}^{(1)} &= \tfrac{A_H k}{4 c \hbar G_M},\\
        \mathscr{S}^{(2)} &= -\tfrac{A_H \pi c^2 \delta \alpha'}{r_0^2} =  - 4\pi^2 c^2 \delta \alpha',
    \end{aligned}
\end{equation}
where $\mathscr{S}^{(1)}$ is the leading order in the entropy and $\mathscr{S}^{(2)}$ is the next to leading order, all higher orders vanish. We note that for Carroll Schwarzschild black hole $\delta = 0$, and we recover the correct result.

 We now calculate the entropy directly from the Carrollian expansion actions. Since black holes can not exist in the electric theory (but will be useful as we will use the constraints derived by the electric action), we use the higher orders beginning with the magnetic action (next-to-leading order) which is given by
\begin{equation}
\begin{aligned}
    S_{(2)}=- \int d^4x & e\big\{-\Bar{R} + \beta'\big[( K^2-K_{\alpha\beta}K^{\alpha\beta})( K^2-K_{\mu\nu}K^{\mu\nu}  + 4\pounds_{\boldsymbol{v}}K)-4(\pounds_{\boldsymbol{v}}K)^2 \big] + \alpha' \big[  -2\bar{R}{}^{\lambda \nu} \nabla_{\mu}(V^{\mu}K_{\lambda \nu}) \\ & - \nabla_{\mu}(V^{\mu}K_{\alpha\beta})K^{\rho \beta} B^{\alpha}{}_{\rho} - \tfrac{1}{2} K^{\rho \alpha} B^{\beta}{}_{ \rho}V^{\mu}\nabla_{\mu}K_{\alpha\beta}  + 2\bar{R}_{\lambda\nu}K^{\lambda \nu} K
    + K^{\lambda \nu} K K^{\alpha}_{\lambda} B_{\nu \alpha}\\ & - \Pi^{\nu\alpha}\nabla_{\mu}K^{\mu}_{\alpha}\nabla_{\rho}K^{\rho}_{\nu} + 2\Pi^{\nu\alpha}\nabla_{\mu}K^{\mu}_{\alpha}\nabla_{\nu}K + V^{\lambda} \nabla_{\mu}K^{\mu}_{\alpha} K^{\rho \alpha} B_{\lambda \rho} 
    - \Pi^{\nu\alpha}\nabla_{\nu}K\nabla_{\alpha}K \\ & -V^{\lambda}\nabla_{\alpha}KK^{\alpha \epsilon} B_{\lambda \epsilon}  - 2V^{\nu}V^{\alpha} \nabla_{\mu}(B_{\alpha}{}^{\mu})\nabla_{\nu}K
    - V^{\alpha}K^{\sigma \lambda} B_{\sigma\alpha} V^{\rho} K^{\beta}_{\lambda} B_{\rho \beta} \\ & + 2 V^{\lambda} \nabla_{\mu}((dT)_{\lambda}{}^{\mu}) K^{\alpha\beta}K_{\alpha\beta}\big] \big\},
    \end{aligned}
\end{equation}
where $e$ is the determinant of the Carrollian metric, and $K_{\mu\nu}$ is the extrinsic curvature after the truncation introduced in \cite{Hansen:2021fxi}. Using the Carroll version of Wald's formula i.e. equation (183) in \cite{Ecker:2023uwm}, and the Hamiltonian constraint derived from the electric action, namely,
\begin{equation}\label{constr}
    \begin{aligned}
         & \ \ \ \ \ K^2-  K^{\mu \nu} K_{\mu \nu} - \alpha' \big(\Pi^{\nu\alpha}\Pi^{\lambda\beta} \nabla_{\mu}(V^{\mu}K_{\alpha \beta})\nabla_{\rho}(V^{\rho}K_{\nu \lambda}) -2 K^{\alpha\beta}K\nabla_{\mu}(V^{\mu}K_{\alpha \beta}) + K^{\lambda \nu}K_{\lambda \nu} K^2 
   \\ &\feq - V^{\mu}V^{\nu}\nabla_{\mu}K\nabla_{\nu}K + 2 K_{\alpha\beta} K^{\alpha \beta} V^{\nu}\nabla_{\nu}K-(K^{\mu\nu}K_{\mu\nu})^2\big) = 0.
    \end{aligned}
\end{equation}
we get the entropy
\begin{equation}
    \mathscr{S}^{(1)} = \tfrac{A_H k}{4 c G_M},
\end{equation}
which is the same as the first order entropy in \eqref{entropy}. The third order Lagrangian in the Carroll expansion is given by
\begin{equation}
\begin{aligned}
    S_{(3)}=  \int d^4x e& \bigg(\tfrac{1}{4} B^{\mu\nu}B_{\mu \nu} - \alpha'\big(  \tfrac{1}{2} \mathcal{K}^{\alpha\beta}\mathcal{K}_{\alpha\beta}B^{\mu\nu}B_{\mu\nu} - \Pi^{\nu\alpha}\nabla_{\alpha}\mathcal{K}\nabla_{\rho}(B_{\nu}{}^{\rho}) - \tfrac{1}{4} V^{\sigma} \mathcal{K}^{\nu\rho} B_{\sigma\rho} \nabla_{\mu}(B_{\nu}{}^{\mu})
      \\ & - \tfrac{1}{2} V^{\lambda} V^{\alpha} \nabla_{\mu}(B_{\alpha}{}^{\mu})\nabla_{\nu}(B_{\lambda}{}^{\nu}) + \tfrac{3}{2}\bar{R}{}^{\alpha\lambda}\mathcal{K}^{\beta}_{\lambda}B_{\alpha\beta}
     + \bar{R}^{\mu\nu}\bar{R}_{\mu\nu} + \tfrac{1}{4} \mathcal{K}_{\beta\lambda}\mathcal{K}^{\rho\lambda}B^{\nu\beta}B_{\nu\rho} + \tfrac{1}{4}\mathcal{K}^{\beta\lambda}\mathcal{K}^{\alpha\rho}B_{\alpha\beta}B_{\lambda\rho}  \big) \\& + \beta' \big( 
 -\mathcal{K}^2\bar{R}  - \mathcal{K}^2\nabla_{\mu}(V^{\lambda}B_{\lambda}^{\ \mu}) + \mathcal{K}_{\mu\nu}\mathcal{K}^{\mu\nu}\bar{R} + \mathcal{K}_{\mu\nu}\mathcal{K}^{\mu\nu}\nabla_{\rho}(V^{\lambda}B_{\lambda}^{\ \rho}) + 2\bar{R}\nabla_{\mu}(V^{\mu}\mathcal{K}) 
        \\& + 2\nabla_{\mu}(V^{\mu}\mathcal{K})\nabla_{\nu}(V^{\lambda}B_{\lambda}^{\ \nu})\big)\bigg),
 \end{aligned}
\end{equation}
Using the Carroll version of Wald's formula and the constraint \eqref{constr}, we get the entropy
\begin{equation}
    \mathscr{S}^{(2)} =  -\tfrac{A_H \pi c^2 \delta \alpha'}{r_0^2},
\end{equation}
which is the same as the next-to-leading order in \eqref{entropy}. By repeating this for higher orders, we see that all higher orders of entropy are zero if the Carroll expansion is truncated as described in \cite{Hansen:2021fxi}.

To sum up, in both cases, the first order of the entropy is given by
\begin{equation}
    \mathscr{S}^{(1)}= \tfrac{A_H k}{4 \hbar c G_M},
\end{equation}
similar to the entropy for Carroll Schwarzschild black hole \cite{Ecker:2023uwm}. The next-to-leading order is given by
\begin{equation}
   \mathscr{S}^{(2)}= -\tfrac{A_H \pi c^2 \delta \alpha'}{r_0^2} = - 4\pi^2 c^2 \delta \alpha'.
\end{equation}
It is evident that it is computationally easier to derive Carroll black holes' entropies by taking the limits of the Lorentzian entropies especially when a closed form is known as the case presented. However, it is useful to know that direct computations from the Carrollian actions using the modified Wald's formula (without referring to a Lorentzian theory) gives the right answer since not all Carrollian theories can be derived from Lorentzian ones. Notice that in the strict Carroll limit $c=0$, the entropy diverges.

 Now we calculate the temperature. Since the temperature is independent of the Bach parameter \cite{Svarc:2018coe}, and since the relativistic temperature is equal to the Carroll temperature in the case of Schwarzschild black hole (using $G_M$ instead of $G$) \cite{Ecker:2023uwm}, then the temperature for Carroll Schwarzschild-Bach black holes is 
\begin{equation}
    T = \tfrac{\hbar c}{8 \pi k M G_M} = \tfrac{\hbar c}{4 k \pi r_0},
\end{equation}
where $M$ is the mass of the black hole. The energy is computed using the first law of thermodynamics to be
\begin{equation}
    \begin{aligned}
        E^{(1)} &= \tfrac{A_H}{32 \pi M G_M^2} = \tfrac{ r_0}{2 G_M},\\
        E^{(2)} &= - \tfrac{2\pi \hbar \delta \alpha' c^3}{k r_0}, 
    \end{aligned}
\end{equation}
where $E^{(1)}$ is the leading order in the expansion for the energy and $E^{(2)}$ is the next-to-leading order. An alternative way to get these results is by using Smarr-Gibbs-Duhem equation \cite{Kubiznak:2016qmn}.

 A problem we would like to address is that in the strict limit the entropy diverges while the temperature goes to zero. This behavior is, in fact, expected since the radius of the black hole is constant (nothing can penetrate the Carroll extremal surface due to the structural singularity \cite{Ecker:2023uwm}, and a Carroll black hole does not emit Hawking radiation \cite{Aggarwal:2024gfb}). Therefore, it can be considered as an incompressible thermodynamical system (in clear contrast to its Lorentzian counterpart). It is known that infinite entropy with zero temperature is consistent and was discussed in \cite{2021arXiv211110315B} and \cite{2013JSMTE..04..010B}. The interpretation is that the system in this limit develops an infinite number of microstates, each of which has an infinitesimal probability, i.e. there are infinitely many ways to form a Carrollian black hole. Here we demonstrate its consistency by calculating the divergent specific heat:
\begin{equation}
    C = -\tfrac{2\pi k r_0^2}{\hbar c G_M},
\end{equation}
which follows from the usual definition for specific heat. This means that (in our case of an incompressible substance) a Carroll black hole can reserve infinite amount of thermal energy in a fixed volume, and for this to be achieved, there must be an infinite number of microstates to accommodate such energy (which aligns with the divergence of the entropy). We also notice that the specific heat is always negative similar to the case of a the Lorentzian Schwarzschild black hole, however, it can not be interpreted as instability in the black hole since Carroll black holes do not radiate as mentioned above.

\section{Carroll Schwarzschild-(A)dS black holes}\label{section 3}
\subsection{Geodesics}\label{sec 4}
In this section, we discuss the Carrollian version of Schwarzschild-(A)dS black hole. In Sec. \ref{sec 2}, we calculated the geodesics and deflection angle formulae with a general function $f(r)$, here we use said formulae in the Carroll Schwarzschild-(A)dS case, i.e, $f(r)=h(r)=1-\frac{2M}{r}- \frac{\Lambda}{3}r^2$. The metric is 
\begin{equation}
    ds^2 = -\left(1-\frac{2M}{r}- \frac{\Lambda}{3}r^2\right)dt^2+\frac{dr^2}{1-\frac{2M}{r}- \frac{\Lambda}{3}r^2} +r^2d\Omega^2.
\end{equation}
The Carrollian black hole is given by
\begin{equation}\label{Carroll (A)dS Schwarzschild Back black hole}
    \begin{aligned}
 h_{\mu\nu}dx^{\mu}dx^{\nu} =  \frac{dr^2}{1-\frac{2M}{r}- \frac{\Lambda}{3}r^2} +r^2d\Omega^2, & \ \ \
 E_{\mu}dx^{\mu} = -\left(1-\frac{2M}{r}- \frac{\Lambda}{3}r^2\right)dt, &
       v^{\mu}\partial_{\mu} = \frac{-1}{1-\frac{2M}{r}- \frac{\Lambda}{3}r^2}\partial_t.
    \end{aligned}
\end{equation}

\noindent Following the same calculations as in Sec. \ref{sec 2}, we obtain the geodesic equation
\begin{equation}
    \frac{\dot{r}^2}{2}+V^{\textrm{eff}}(r)=E,
\end{equation}
where the effective potential in this case reads $V^{\textrm{eff}}(r)=\frac{l^2}{2r^2}(1-\frac{2M}{r}-\frac{\Lambda r^2}{3}) + (\frac{2M}{r}+\frac{\Lambda r^2}{3})E$. The circularity conditions are
\begin{equation}
    \begin{aligned}
        \left(1-\frac{2M}{r}- \frac{\Lambda}{3}r^2\right)\left(\frac{1}{b^2}-\frac{1}{r^2}\right)l^2=0, \\
        \left(\frac{1}{b^2}-\frac{1}{r^2}\right)l^2\left(\frac{2M}{r^2}-\frac{2\Lambda r}{3}\right)+ \frac{l^2}{r^2}\left(1-\frac{2M}{r}- \frac{\Lambda}{3}r^2\right)=0.
    \end{aligned}
\end{equation}

As before, they are satisfied at the extremal surface, i.e, at $r=r_0$ with the condition
\begin{equation}
    b=r_0.
\end{equation}
 
 Using the same argument as in Sec. \ref{sec 2}, there is no orbit motion outside the extremal surface. After calculating the second derivative of the effective potential $V^{\textrm{eff}}(r)$, and evaluating at $r_0$, we obtain
\begin{equation}
    \frac{d^2V^{\textrm{eff}}}{dr^2}(r_0)=-\frac{2l^2}{r_0^3}\left(\frac{2M}{r_0^2}-\frac{2\Lambda r_0}{3}\right).
\end{equation}

We have the following three cases:
\begin{enumerate} \itemsep0em 
\item $\Lambda<\frac{3M}{r_0^3}$: $ \tfrac{d^2V^{\textrm{eff}}}{dr^2}(r_0)$ is negative, and there are no stable orbits. Since $M, r_0$ are positive, therefore, if we have a negative cosmological constant, we cannot have stable orbits. 
\item $\Lambda > \frac{3M}{r_0^3}$:  $ \tfrac{d^2V^{\textrm{eff}}}{dr^2}(r_0)$ is positive and there are stable orbits.
\item $\Lambda = \frac{3M}{r_0^3}$, then $\tfrac{d^2V^{\textrm{eff}}}{dr^2}(r_0) = 0$ and the second derivative test is inconclusive.
\end{enumerate}
From this we conclude that stable orbits can exist on the extremal surface of Carroll Schwarzschild-(A)dS black holes, and the requirement for that to happen depends on the mass as well as the value of the cosmological constant.

 We now move on to calculating the deflection angle of a geodesic particle. We begin by the equation
\begin{equation}
    \frac{d \phi}{dr}=\frac{b}{r\sqrt{r^2-b^2}} \frac{1}{\sqrt{1-\frac{2M}{r}-\frac{\Lambda r^2}{3}}}.
\end{equation}
Changing variables to $u=\frac{1}{r}$ and integrating,
the deflection angle is now given by
\begin{equation}
    \phi = -\int_0^{1/b} \frac{b}{\sqrt{1-b^2u^2}} \frac{1}{\sqrt{1-2Mu-\frac{\Lambda}{3u^2}}} du.
\end{equation}
This integral is not solvable analytically so we explore the approximation when $\Lambda$ is small. For this purpose, we call the integrand $\psi$, i.e,
\begin{equation}
    \psi = - \frac{b}{\sqrt{1-b^2u^2}} \frac{1}{\sqrt{1-2Mu-\frac{\Lambda}{3u^2}}}.
\end{equation}
The approximation to the linear order in $\Lambda$ is then given by
\begin{equation}
   \phi = -2\int_0^{1/b} \frac{b}{\sqrt{1-b^2u^2}} \frac{1}{\sqrt{1-2M u}} du +2 \Lambda \int_0^{1/b} \frac{\partial \psi}{\partial \Lambda}|_{\Lambda=0} \  du.
\end{equation}
The first integral is the same as that of Schwarzschild and the second integral will serve as a correction due to the presence of the cosmological constant. The second integral diverges in the upper limit, so we introduce a regularization parameter $a$, integrate from 0 to $1/b - a$, then expand as a series of powers of $a$, keeping only the leading term. We will see that by appropriately fixing $a$, we get a finite deflection angle. The result of the integrals is
\begin{equation}
    \phi = \tfrac{4}{\sqrt{1+x}}\bigg( 
K(\tfrac{2x}{1+x}) - F(\tfrac{\pi}{4},\tfrac{2x}{1+x}) \bigg) + 2\Lambda \bigg( \sqrt{\tfrac{b}{2}} \tfrac{1}{\sqrt{a}\sqrt{1-x}} -b  \bigg),
\end{equation}
where we denoted $x=\tfrac{2M}{b}$. Now we examine two limits : $x \to 0$ where the mass goes to zero, i.e., the limit where the black hole is massless (does not exist), and limit where the incoming particle is tangential to the extremal horizon. Since the radius of the extremal horizon satisfies $r_H(1+\tfrac{\Lambda r_H^2}{3}) = 2M$, the limit should be $x \to 1 + \tfrac{\Lambda r_H^2}{3}$, however, we can shift $x$ by $\tfrac{-\Lambda r_H^2}{3}$ and use the limit $x \to 1$. In the first limit, the expansion to the leading order in $x$ is
\begin{equation}\label{defl}
    \phi = \pi - 2\Lambda b  + \sqrt{2b} \tfrac{\Lambda}{\sqrt{a}}+ (1 + \sqrt{\tfrac{b}{2}}\tfrac{\Lambda}{2 \sqrt{a}})x. 
\end{equation}
Since if $x=0$ we expect $\phi = \pi$ in the limits $\Lambda \to 0$ and $a \to 0$, we require that $\tfrac{\Lambda}{\sqrt{a}}$ goes to zero in this limit, i.e. $\Lambda$ goes to zero at least as fast as $a$. Thus, since we have the freedom to choose $a$, we choose $a = |\Lambda|$, and so $\tfrac{\Lambda}{\sqrt{a}} = \pm \sqrt{|\Lambda|}$. We use this to examine the other limit $x \to 1$. In that limit, to the first order in $\Lambda$, the expansion is
\begin{equation}
    \phi = -\sqrt{2}\textrm{ln}(1-x + \tfrac{|\Lambda|r_H^2}{3}) + \Delta \pm \tfrac{\sqrt{2 b |\Lambda|}}{\sqrt{1-x + \tfrac{|\Lambda|r_H^2}{3}}} - 2 \Lambda b,
\end{equation}
where $\Delta = \sqrt{2}\textrm{ln}(\tfrac{32}{(1+\sqrt{2})^2})$. As a consistency check, the expansions in the two limits reduce to the expansion in \cite{Ciambelli:2023tzb} if we set $\Lambda =0$.

\begin{figure}[ht]
    \centering
    \includegraphics{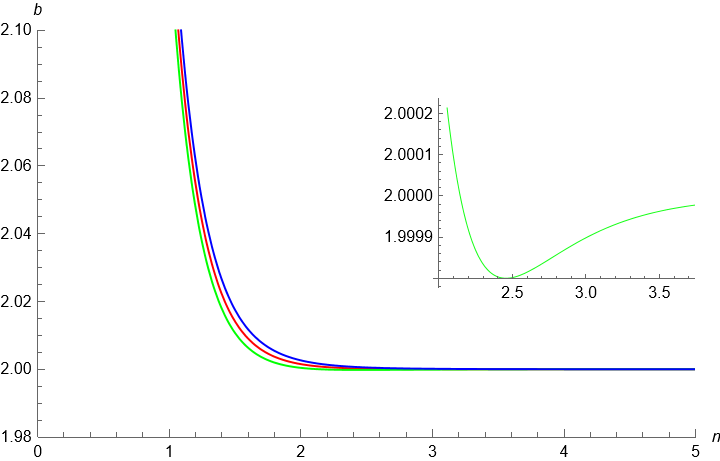}
    \caption{This graph represents the impact parameter as a function of the winding number in the cases of Carroll Schwarzschild (red), Carroll Schwarzschild-dS (green) and Carroll Schwarzschild-AdS (blue) with $M=1$ and $\sqrt{|\Lambda|}=0.02$. For a given value of the impact parameter $b>2$, we see that Carroll Schwarzschild-dS winds less than Carroll Schwarzschild which winds less than Carroll Schwarzschild-AdS, which is consistent with the conclusions above. For $b=2$, the winding number can be arbitrarily high. This coincides with the result in \cite{Ciambelli:2023tzb} in the case of Carroll Schwarzschild.  And although $b=2$ corresponds to a tangential particle in the cases of Carroll Schwarzschild and Carroll Schwarzschild-AdS, it is not the case for Schwarzschild-dS where a tangential particle has $b=1.99998$. Thus, the graph is not monotonically decreasing like the others instead it has a local minimum corresponding to a tangential particle, then $\phi$ goes up and stabilizes outside the extremal surface. This means that a tangential particle winds 2.48 times while the arbitrary high winding number occurs just outside the extremal surface. This is a result of the fact that at these particular values of the parameters, we have two horizons i.e. it is expected to have two winding numbers (one around each horizon) for some values of the impact parameter. On the other hand, for large impact parameter, we have low winding number in all cases as expected.}
    \label{fig:1}
\end{figure}

As in Schwarzschild's case, a geodesic particle will wind around the black hole with a winding number depending on the impact parameter. When the impact parameter is close enough to the radius of the extremal surface, we get a large deflection angle $\phi \geq 2\pi$, then we can define the number of windings $n$ undergone by the particle around the extremal surface as $\phi = 2 n \pi $, where $n \geq 1$. Using that in \eqref{defl}, we get a transcendental equation whose solution gives the impact parameter as a function of the number of revolutions in this approximation. This transcendental equation is 
\begin{equation}
    1-x + \tfrac{|\Lambda|r_H^2}{3} = \exp{\bigg({(2n \pi - \Delta)/\sqrt{2} \pm \tfrac{2 \sqrt{M |\Lambda|}}{\sqrt{x(1-x^2 + \tfrac{|\Lambda|r_H^2}{3})}} + \tfrac{4M \Lambda}{x}}\bigg)}.
\end{equation}
Such equation has no analytical solution, so we use perturbation theory to find an approximate solution to the linear order of $\Lambda$ (which is consistent since we assumed that $|\Lambda| \ll 1$). The solution, after substituting with $x = \tfrac{2M}{b}$ and neglecting higher orders in $\Lambda$, is given by
\begin{equation}
    b(n) = \frac{2M}{1- e^{-(2n \pi - \Delta)/\sqrt{2}} \pm 2 \sqrt{M |\Lambda| }\sqrt{\tfrac{e^{-(2n \pi - \Delta)/\sqrt{2}}}{1 - e^{-(2n \pi - \Delta)/\sqrt{2}}}}},
\end{equation}
where the plus sign is in the case of dS space and the minus sign is for the AdS case. We notice that for the same impact parameter, the number of windings is less than Carroll Schwarzschild in the case of a positive cosmological constant but more than Carroll Schwarzschild in the case of a negative cosmological constant. This behaviour is visualized in Fig. \ref{fig:1}.

\subsection{Thermodynamics}\label{sec 5}
Now we will examine the thermodynamics of Carroll Schwarzschild-(A)dS black holes by calculating the entropy, temperature, and energy of such black holes. Since the entropy is independent of the cosmological constant \cite{Svarc:2018coe} one can see that 
the entropy for Carroll Schwarzschild-(A)dS is equal to the entropy for Carroll Schwarzschild black hole, namely,
\begin{equation}\label{eq}
    \mathscr{S}=\tfrac{A_H k}{4 c G_M \hbar}.
\end{equation}
If $0< 9\Lambda M^2 < 1$, there are two horizons in the Lorentzian case which translates into two extremal surfaces in the Carrollian case, namely, a black hole extremal surface with entropy $\mathscr{S}_H$ equal to \eqref{eq}, and a cosmological extremal surface with entropy $\mathscr{S}_C = \tfrac{A_c k}{4 c G_M \hbar}$, where $A_c$ is the area of the cosmological extremal surface. The total entropy is
\begin{equation}
     \mathscr{S}=\tfrac{(A_H+A_c )k}{4 c G_M \hbar}.
\end{equation}

 To calculate the temperature for the Carroll black hole, we first do that for the Lorentzian black hole. we first derive the surface gravity at the horizon. Surface gravity can be computed from the formula $\kappa_H =\tfrac{c^2 f'(r_H)}{2}$ (in standard units), where in this case $f(r) = 1- \tfrac{2M}{r} - \tfrac{\Lambda r^2}{3}$. In standard units the surface gravity is given by
\begin{equation}
    \kappa_H = \tfrac{c^2}{2}\big(\tfrac{1}{ r_H} - \Lambda r_H\big).
\end{equation}
The temperature is then
\begin{equation}
    T = \tfrac{\hbar c }{4\pi k} \big( \tfrac{1}{ r_H} - \Lambda r_H \big).
\end{equation}
which is of the same form as in the Carrollian case (since as mentioned before, the temperature does not change on taking the Carroll limit).

 If $0<9\Lambda M^2<1$, i.e, we have a cosmological extremal surface, its surface gravity and the temperature would be similar to the black hole's formula with $r_H$ is replaced by the radius of the cosmological extremal surface $r_c$, however, we have an effective temperature of \cite{Bhattacharya:2015mja,Urano:2009xn,Bhattacharya:2013tq}
\begin{equation}\label{effective temperature}
    T_{\textrm{eff}}= \tfrac{\hbar \kappa_H \kappa_c}{4 \pi c k(\kappa_H + \kappa_c)},
\end{equation}
where $\kappa_c$ is the surface gravity on the cosmological extremal surface. Notice that $T_{\mathrm{eff}}$ goes to zero even faster than the temperature of each horizon separately.

 The energy is calculated through the first law of thermodynamics to be
\begin{equation}
    E = \tfrac{1}{2G_M}\big(r_H - \tfrac{\Lambda}{3}r_H^3\big),
\end{equation}
which also agrees with the result from Smarr-Gibbs-Duhem formula in \cite{Kubiznak:2016qmn}. If $0<9 \Lambda M^2<1$, the cosmological extremal surface will have energy given by a formula similar to $A_H$ replaced by $A_c$, and the total energy would be the sum of the two. Notice that the energy in this case also goes to zero faster than in the case of Carroll Schwarzschild and Carroll Schwarzschild-Bach black holes.

 As in Sec. \ref{sec.3}, the entropy diverges, while the temperature goes to zero in the strict limit. Thus, we have a divergent specific heat given by
\begin{equation} \label{specific heat}
    C = \tfrac{2\pi k r^2_H}{\hbar c G_M} (\tfrac{\Lambda r^2_H -1}{\Lambda r^2_H + 1}). 
\end{equation}
The specific heat derived in this case diverges at the strict limit as expected; however, it is not always negative. For a non zero value of $c$, it can be positive, negative, or zero. This leads to a classification of Carroll black holes similar to the classification for Lorentzian black holes explored in \cite{Kubiznak:2014zwa} and \cite{Kubiznak:2016qmn} but with different interpretations since negative specific heat can no longer be interpreted as instability due to the lack of Hawking radiation \cite{Aggarwal:2024gfb} and incompressibility. We will consider this classification in future work.

\section{Carroll Schwarzschild-Bach-(A)dS black holes}\label{section 4}
\subsection{Geodesics}\label{sec 6}
Finally, we define the Carroll Schwarzschild-Bach-(A)dS black hole. The Lorentzian metric is given by
\begin{equation}
    ds^2 = -h(r)dt^2 + \frac{dr^2}{f(r)} + r^2d\Omega^2,
\end{equation}
where $f(r)$ and $h(r)$ are functions defined by their expansions near the horizon. We define the Carrollian black hole counterpart similar to the previous sections as
\begin{equation}
    \begin{aligned}
        h_{\mu\nu}dx^{\mu}dx^{\nu}= \frac{dr^2}{f(r)} + r^2d\Omega^2, & \ \ \ E_{\mu}dx^{\mu}= -h(r)dt, & v^{\mu}\partial_{\mu}=\frac{-1}{h(r)}\partial_t.
    \end{aligned}
\end{equation}

Following the same procedure as in Sec. \ref{sec 2}, we get the same form of the geodesic equation \eqref{geodesic equation} with the same form of the effective potential, and the same circularity conditions as \eqref{circularity conditions}, where they are satisfied at $r=r_0$ with the condition $b=r_0$ i.e. the forms for these formulae do not change if we add a cosmological constant. In a similar way to the Schwarzschild-Bach case, the Schwarzschild-Bach-(A)dS solution is known only near the horizon. Thus, we expand $f(r)$ into powers of $r-r_0$ as follows:
\begin{equation}
    f(r)= f_1 (r-r_0) + f_2 (r-r_0)^2+\cdots,
\end{equation}
where $f_1=\frac{1-\Lambda r_0^2+\delta}{r_0}$ and $f_2 = \frac{1}{r_0^2(1-\Lambda r_0^2+\delta)}[1-\Lambda r_0^2 + \delta (3-\Lambda r_0^2 - \tfrac{3}{8k}r_0^2)+2\delta^2]$ \cite{Svarc:2018coe,Pravdova:2023nbo}. One can easily see that it simplifies to the formulae in Sec. \ref{sec 2} when setting $\Lambda =0$ and of the formulae in Sec. \ref{sec 4} when setting $\delta = 0$.

 The stability condition is given by calculating the second derivative of the effective potential and evaluating it at the extremal surface. The result is
\begin{equation}   \frac{d^2V^{\textrm{eff}}}{dr^2}(r_0) = -\frac{2l^2}{r_0^4}(1-\Lambda r_0^2 + \delta).
\end{equation}
We have three cases:
\begin{enumerate} \itemsep0em 
 \item $1+\delta > \Lambda r_0^2$: $ \tfrac{d^2V^{\textrm{eff}}}{dr^2}(r_0)$ is negative and there are no stable orbits.
 \item $1+\delta < \Lambda r_0^2$: $ \tfrac{d^2V^{\textrm{eff}}}{dr^2}(r_0)$ is positive and there are stable orbits.
 \item $1+\delta = \Lambda r_0^2$: $\tfrac{d^2V^{\textrm{eff}}}{dr^2}(r_0) = 0$ and the second derivative test is inconclusive.
\end{enumerate}
\noindent This means that stable orbits can be generated by adjusting two parameters ($\delta$ and $\Lambda$). This means that there are more stable orbits when compared to those of Carroll Schwarzschild-Bach or Carroll Schwarzschild-(A)dS black holes.

 As done in the previous sections, we now calculate the deflection angle for a geodesic particle. We use the formula
\begin{equation}
    \frac{d\phi}{dr} = \frac{b}{r\sqrt{r^2-b^2}}\bigg( \frac{1}{\sqrt{f_1 (r-r_0)}} - \frac{f_2}{2f_1^{3/2}}\sqrt{r-r_0} \bigg),
\end{equation}
 changing variables to $u=\frac{1}{r}$ and $u_0 = \frac{1}{r_0}$, then integrating, 
the deflection angle is
\begin{equation}
    \begin{aligned}
       \phi &= \frac{1}{(1-\frac{\Lambda}{u_0^2} +\delta)^{3/2} \sqrt{\frac{b+\frac{1}{u}}{b}} \sqrt{1-b^2 u^2} \sqrt{u_0-u}} \bigg(b u^{3/2} \sqrt{\frac{1-\frac{u_0}{u}}{b u_0+1}} \bigg(\sqrt{\frac{b+\frac{1}{u}}{b}} \sqrt{1-\frac{1}{b^2 u^2}} (2 (1-\frac{\Lambda}{u_0^2} +\delta)  \\& + \frac{1}{(1-\tfrac{\Lambda}{u_0^2} +\delta)}[1-\tfrac{\Lambda}{u_0^2} + \delta (3-\tfrac{\Lambda}{u_0^2} - \tfrac{3}{8u_0^2k})+2\delta^2]) \Pi \bigg(2;\sin ^{-1}\bigg(\frac{\sqrt{\frac{b+\frac{1}{u}}{b}}}{\sqrt{2}}\bigg)|\frac{2 b u_0}{b u_0+1}\bigg)
        +\frac{u_0}{(1-\tfrac{\Lambda}{u_0^2} +\delta)} \\& \times [1-\tfrac{\Lambda}{u_0^2} + \delta (3-\tfrac{\Lambda}{u_0^2} - \tfrac{3}{8u_0^2k})+2\delta^2] \bigg(b+\frac{1}{u}\bigg) \sqrt{1-\frac{1}{b u}} F\bigg(\sin ^{-1}\bigg(\frac{\sqrt{\frac{b+\frac{1}{u}}{b}}}{\sqrt{2}}\bigg)|\frac{2 b u_0}{b u+1}\bigg)\bigg)\bigg).
    \end{aligned}
\end{equation}
Similar to the results in Sec. \ref{sec 2}, the deflection angle diverges at $u=1/b$, so we introduce the regularization parameter $a$ and evaluate it at $u= 1/b - a$. The difference in this case is that $f_1$ and $f_2$ can be infinite when taking $\Lambda \to \infty$. For any finite value for $\Lambda$, the deflection angle diverges as in Sec. \ref{sec 2}, and we come to the same conclusion that adding higher-derivative terms implies that particles close enough to the extremal surface will wind an infinite number of times. In the limit $\Lambda \to \infty$ such that it grows faster than $a^{1/3}$, then $\phi$ goes to zero, however, such spacetimes cannot accommodate black holes \cite{Aldrovandi:2001vx}, thus we neglect this possibility.

\subsection{Thermodynamics}\label{sec 7}
Now, we examine the thermodynamics of Carroll Schwarzschild-Bach-(A)dS black hole. We calculate the entropy, temperature, and energy and comparing the results with the previous black holes considered. In this section, we have to differentiate between two cases:
\begin{enumerate}
 \item  The metric is a solution for theory (2,4) with an added cosmological constant.
  \item  The metric is a solution for the theory (4,2) without an added cosmological constant (since (4,2) has an emergent cosmological constant $\Lambda = \tfrac{-1}{10\beta'}$ \cite{Tadros:2023teq}).
\end{enumerate}
We can see from \cite{Svarc:2018coe} or using the Carrollian version of Wald's formula that the two cases give similar formulae for the entropy up to a power of $c$ in the next-to-leading order (the first with a general value for the cosmological constant and the second with $\Lambda = \tfrac{-1}{10\beta'}$): In the case of (2,4), the entropy is
\begin{equation}
    \begin{aligned}
        \mathscr{S}^{(1)}&=\tfrac{A_H k}{4 c G_M \hbar},\\
        \mathscr{S}^{(2)}&= c^2 \alpha' \pi \big( \tfrac{A_H \Lambda}{3} - 4\pi \delta \big),\\ 
        \mathscr{S}^{(3)} &= 2A_H \pi \Lambda c^4 \beta',
    \end{aligned}
\end{equation}
where $\mathscr{S}^{(1)}$ is the leading order, $\mathscr{S}^{(2)}$ is the next-to-leading order and $\mathscr{S}^{(3)}$ is the next-to-next-to-leading order with all higher orders are zero. In the case when $0<9\Lambda M^2<1$, we have two extremal surfaces as discussed before and the total entropy is the sum of the two entropies. In the case of (4,2), the entropy is
\begin{equation}\label{entropy 2}
    \begin{aligned}
        \mathscr{S}^{(1)}&=\tfrac{A_H k}{4 c G_M \hbar},\\
        \mathscr{S}^{(2)} &= \tfrac{-A_H \pi c^2}{5},\\
        \mathscr{S}^{(3)} &= -c^4 \pi \alpha' \big(\tfrac{A_H}{30\beta'} + 4\pi \delta \big)
    \end{aligned}
\end{equation}
Note that in this case, the next-to-leading order goes to zero slower than in the (2,4) case. Also, since this theory does not permit a cosmological extremal surface, equations \eqref{entropy 2} are valid without restrictions.

 The temperature is independent of the Bach parameter \cite{Svarc:2018coe}, thus, the temperature is the same as the Carroll Schwarzschild-(A)dS case. Namely, for (2,4) we have
\begin{equation}
     T = \tfrac{\hbar c }{4\pi k} \big( \tfrac{1}{ r_0} - \Lambda r_0 \big).
\end{equation}
If there is a cosmological extremal surface, we have an effective energy equal to
\begin{equation}
    T_{\textrm{eff}}= \tfrac{\hbar \kappa_H \kappa_c}{4 \pi c k(\kappa_H + \kappa_c)},
\end{equation}
 while in the case of (4,2), the temperature is
\begin{equation}
     T = \tfrac{\hbar c }{4\pi k} \big( \tfrac{1}{ r_0} + \tfrac{r_0}{10\beta'} \big).
\end{equation}

 The energy is again calculated using the first law of thermodynamics (or by Smarr-Gibbs-Duhem formula). In the case of (2,4), we have
\begin{equation}
    \begin{aligned}
        E^{(1)}&= \tfrac{1}{2 G_M}\big(r_0 - \tfrac{\Lambda r_0^3}{3}\big),\\
         E^{(1)}&= \tfrac{2 \pi \hbar c^3 \alpha' \Lambda}{3k}\big(r_0 - \tfrac{\Lambda r_0^3}{3}\big),\\
        E^{(3)} &= \tfrac{4 \pi \hbar c^5 \Lambda \beta'}{k}\big(r_0 - \tfrac{\Lambda r_0^3}{3}\big).
    \end{aligned}
\end{equation}
If $0<9\Lambda M^2<1$, then we have a cosmological extremal surface and the energy for it is given by similar formulae, but replacing $r_H$ and $A_H$ with $r_c$ and $A_c$ respectively. The total energy is the sum of both extremal surfaces' energies. The energy for the (4,2) case is given by
\begin{equation}
    \begin{aligned}
         E^{(1)}&= \tfrac{1}{2 G_M}\big(r_0 + \tfrac{ r_0^3}{30\beta'}\big),\\
         E^{(2)} &= -\tfrac{2 \pi \hbar c^3}{5k}\big(r_0 + \tfrac{ r_0^3}{30\beta'}\big),\\
        E^{(3)} &= -\tfrac{ \pi \hbar c^5 \alpha'}{15 k \beta'}\big(r_0 + \tfrac{ r_0^3}{30\beta'}\big).
    \end{aligned}
\end{equation}

\noindent Again, entropy diverges while the temperature goes to zero. Thus, we calculate a divergent specific heat as follows: For (2,4) the specific heat is 
\begin{equation}
    C =  (\tfrac{2\pi k r^2_0}{\hbar c G_M} - \tfrac{8\pi^2 c^2 \Lambda \alpha'r_0^2}{3} - 16 \pi^2 \Lambda c^4 \beta') (\tfrac{\Lambda r^2_0 -1}{\Lambda r^2_0 + 1}),
\end{equation}
while for (4,2), the specific heat is
\begin{equation}
  C =  (\tfrac{2\pi k r^2_0}{\hbar c G_M} + - \tfrac{8\pi^2 c^4 \Lambda \alpha'r_0^2}{3} + \tfrac{4 c^2}{10}) (\tfrac{ r^2_0 +10\beta'}{r^2_0 - 10\beta'}).  
\end{equation}
In both cases, the specific heat can be positive, negative, or zero. This gives rise to a classification as discussed in the previous sections. Since the classification of Carroll black holes is done in the Carroll limit, and is only concerning the specific heat. And since the only important part for this classification is the divergent part, we can write the specific heat in a more practical way as
\begin{equation}
    C = \tfrac{2\pi k r^2_0}{\hbar c G_M}(\tfrac{\Lambda r^2_0 -1}{\Lambda r^2_0 + 1}) + \textrm{convergent terms},
\end{equation}
for (2,4) theory, and
\begin{equation}
    C = \tfrac{2\pi k r^2_0}{\hbar c G_M}(\tfrac{ r^2_0 +10\beta'}{r^2_0 - 10\beta'}) + \textrm{convergent terms},
\end{equation}
for (4,2) theory, where the convergent terms will not affect the classification so can be dropped out. In both cases, the resulting classification would be more robust since it depends on an additional parameter $\beta'$, and there is the case where there exists a cosmological extremal surface. We delegate such classification to future work.

\section{Conclusions and outlook}\label{sec 8}
In this paper, we introduced Carroll Schwarzschild-(A)dS and Carroll Schwarzschild-Bach-(A)dS black holes and examined the dynamics of geodesic particles in their geometries. We found that in the case of Schwarzschild-(A)dS, in a small cosmological constant approximation, a nearly tangential particle winds around the Carroll extremal surface more than Carroll Schwarzschild in the case of a negative cosmological constant, while it winds less in the case of a positive cosmological constant. The case of Schwarzschild-Bach-(A)dS however is different, a nearly tangential particle has an infinite winding number i.e. it never escapes the vicinity of the extremal surface, and that is true for any value for the cosmological constant. Thus, we interpret that as the effect of adding higher curvature terms in the Lagrangian.

In addition to geodesic particle dynamics, we analyzed the thermodynamics of Carroll black holes. The most notable aspect is the divergence of entropy when the temperature goes to zero, which is also present in Carroll Schwarzschild's case. This means that there are an infinite number of microstates to these black holes, each of which has an infinitesimal probability. This agrees with the fact that the specific heat is calculated to be divergent, i.e. since the black holes have an infinite number of microstates, they can accommodate infinite thermal energy in a finite volume. Similar to the Lorentzian case, this leads to a classification of black holes according to whether the specific heat is positive or negative; this classification is harder in Schwarzschild-Bach-(A)dS due to the presence of an additional parameter (the Bach parameter). However, a notable simplification from the Lorentzian case is that Carroll black holes can be considered as incompressible thermodynamical systems. This arises from the facts that particles can not enter the extremal surface, and Carroll black holes do not radiate. Thus, the specific heat at constant pressure and at constant volume are the same, and the classification is simplified from the Lorentzian case, where we have to take all variables into consideration. 

We also notice that the thermodynamical quantities for the studied black holes are similar to their Lorentzian counterpart except for the order of $c$ and the incompressibility. This can lead to a (thermodynamical) correspondence between Carroll black holes and Lorentzian ones if we consider an ensemble of Carroll black holes parameterized by the volume. The fact that Carroll black holes satisfy the first law of thermodynamics leads to the expectation that the thermodynamical diagrams should be similar but with different interpretations. For example, a negative specific heat can not be interpreted as an instability (since Carroll black holes do not radiate), and no phase transition can happen (due to incompressibility). Such duality can simplify black hole thermodynamics significantly in the case of higher-derivative theories where there are other parameters in the theory, the reason is that in the Carroll case, diagrams have one less parameter but still give the same classification. The exploration of such diagrams and the link to Lorentzian black holes will be a topic for future work. 

Another interesting topic to be considered is the extension of the results in this paper to rotating and charged black holes, and if energy can be extracted from them by the means used in the Lorentzian case.

\section*{Acknowledgements}

The authors would like to thank Luca Ciambelli (Peremiter institute, Canada), Daniel Grumiller (Vienna, Austria), Pavel Krtou\v{s}, David Kubiz\v{n}\'ak, David Heyrovsk\'y (Prague, Czech republic) for stimulating discussions. P.T. and I.K. were supported by Primus grant PRIMUS/23/SCI/005 from Charles University. I.K. also acknowledges financial support from Research Center Grant No. UNCE24/SCI/016 from Charles University.

\end{document}